\documentclass[twocolumn,prl,aps,nobalancelastpage]{revtex4}

\usepackage{graphicx}
\usepackage{bm}
\begin{document}
\title{Shubnikov-de Haas oscillations in YBa$_2$Cu$_4$O$_8$}

\author{A. F. Bangura$^1$, J. D. Fletcher$^1$, A. Carrington$^1$, J. Levallois$^2$, M. Nardone$^2$, B. Vignolle$^2$
, P. J. Heard$^1$, N. Doiron-Leyraud$^3$, D. LeBoeuf$^3$, L. Taillefer$^3$, S. Adachi$^4$, C.
Proust$^2$ and N. E. Hussey$^1$}

\affiliation{$^1$H. H. Wills Physics Laboratory, University of Bristol, Tyndall Avenue, BS8 1TL,
United Kingdom.}

\affiliation{$^2$Laboratoire National des Champs Magn\'{e}tiques Puls\'{e}s, UMR CNRS-UPS-INSA
5147, Toulouse, France.}

\affiliation{$^3$D\'{e}partement de physique and RQMP, Universit\'{e} de Sherbrooke, Sherbrooke,
J1K 2R1, Canada.}

\affiliation{$^4$Superconducting Research Laboratory, International Superconductivity Center,
Shinonome 1-10-13, Tokyo 135, Japan.}

\date{\today}
\begin{abstract}
We report the observation of Shubnikov-de Haas oscillations in the underdoped cuprate superconductor YBa$_2$Cu$_4$O$_8$
(Y124).  For field aligned along the $c$-axis, the frequency of the oscillations is $660\pm 30$~T, which corresponds to
$\sim 2.4$\% of the total area of the first Brillouin zone.  The effective mass of the quasiparticles on this orbit is
measured to be $2.7\pm0.3$ times the free electron mass.  Both the frequency and mass are comparable to those recently
observed for ortho-II YBa$_2$Cu$_3$O$_{6.5}$ (Y123-II). We show that although small Fermi surface pockets may be
expected from band structure calculations in Y123-II, no such pockets are predicted for Y124. Our results therefore
imply that these small pockets are a generic feature of the copper oxide plane in underdoped cuprates.
\end{abstract}
\maketitle

Understanding the doping evolution of the electronic ground state within the superconducting CuO$_2$ planes is central
to the high-$T_c$ problem, yet many outstanding issues remain. Heavily overdoped cuprates, for example, behave in many
ways like conventional, highly anisotropic metals \cite{Proust02, Nakamae03, HusseyACMB03}, yet quantum oscillations, a
characteristic signature of a Fermi-liquid ground state \cite{Shoenberg}, have never been observed. Moreover, it is
usually assumed that as hole doping is decreased and the materials evolve towards the antiferromagnetic insulating
state, the properties depart substantially from those of a conventional Fermi-liquid. The recent observation
\cite{Doiron-leyraudPLLBLBHT07} of the Shubnikov de Haas effect (SdH) in underdoped YBa$_2$Cu$_3$O$_{6.5}$, with
$T_c/T^{\rm max}_c\simeq 0.6$, was therefore a surprising and highly significant discovery.

The frequency of the SdH oscillations in oxygen ordered ortho-II YBa$_2$Cu$_3$O$_{6.5}$ (Y123-II) corresponds to a
Fermi surface pocket with a cross-sectional area (for {\bf H}$\|c$) that is only a small fraction ($\sim$2\%) of the
total Brillouin zone area. The angle dependence of the frequency of this oscillation suggests that this pocket is quasi-two dimensional~\cite{Jaudet07}.
Density functional theory (DFT) band-structure calculations \cite{CarringtonY07,ElfimovSD07}
suggest that these pockets may be formed from a flat band, originating from the CuO chain and BaO layers, which lies
very close to the Fermi level $E_F$. However, there are a number of alternative possibilities such as the formation of
pockets near the nodal points of the DFT CuO$_2$ plane Fermi surface, where the $d$-wave superconducting gap is
minimal, and where the most intense spectral weight is observed by angle resolved photoemission spectroscopy (ARPES)
\cite{DamascelliHS03}. These pockets or \lq Fermi arcs' may occur due to promixity to the Mott insulating state
\cite{Kyung06} or because of reconstruction of the original Fermi surface resulting from the formation of an ordered
phase with broken symmetry \cite{Chakravarty01, LinMillis05, ChenYRZ07}. Clearly, it is imperative to determine whether
the SdH signals are unique to Y123-II or are an intrinsic feature of the electronic structure of the cuprates.

In this Letter we report the observation of SdH oscillations in another underdoped cuprate superconductor
YBa$_2$Cu$_4$O$_8$ (Y124), which has a higher $T_c$ ($\sim$ 80 K) and a higher hole density per planar Cu ($p \sim
0.14$) than Y123-II ($p \sim 0.1$), in pulsed high magnetic fields up to 61 Tesla. The frequency of the oscillations is
comparable with those in Y123-II. This similarity, coupled with the marked differences in the calculated band-structure
of the two materials, strongly points to an origin of the oscillations which is generic to the underdoped cuprates.

\begin{figure}
\includegraphics*[width=0.70\linewidth,clip]{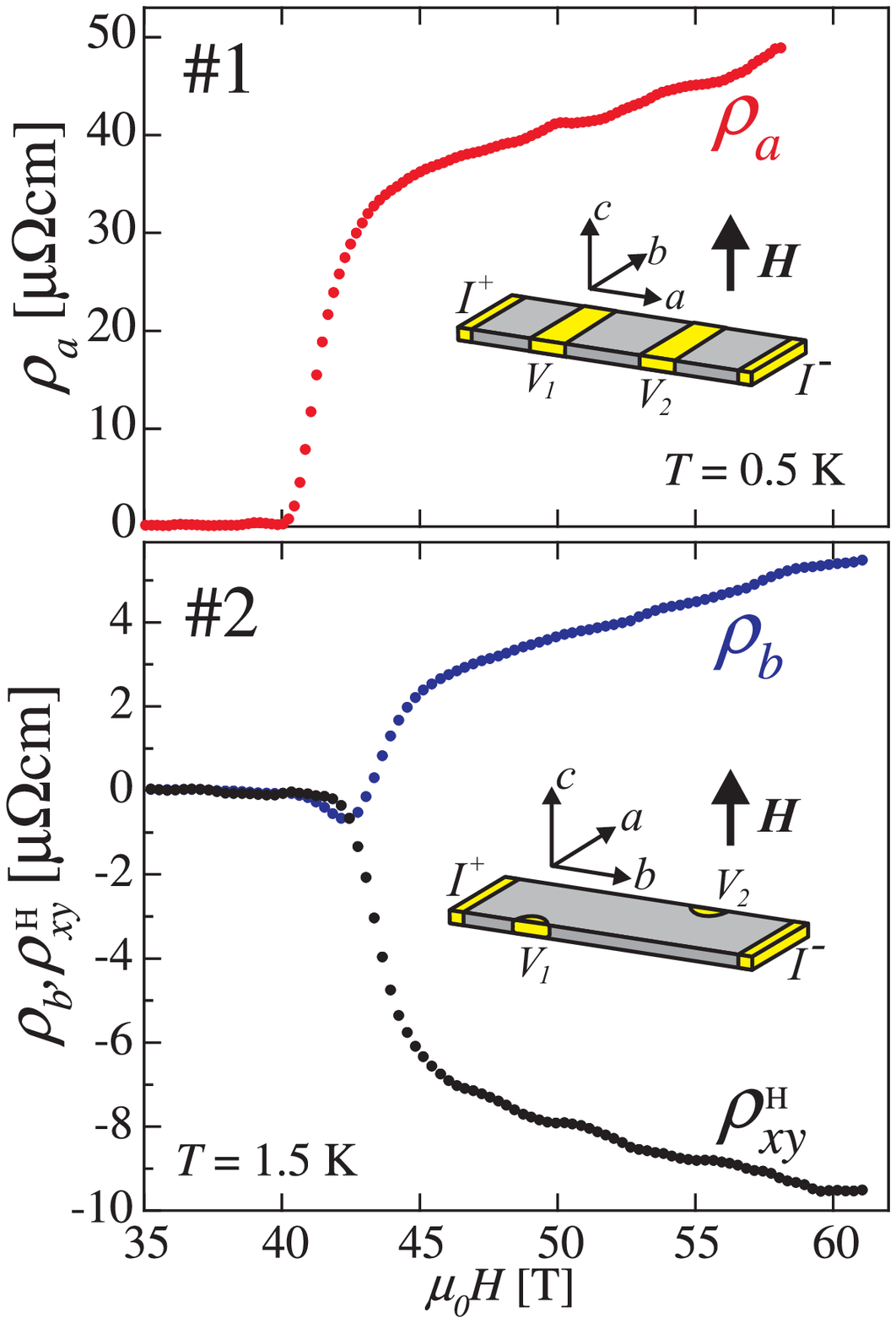}
\caption{(Color online) Longitudinal and Hall data ($\rho_{xx}$, $\rho_{yy}$ and $\rho_{ xy}$) for
two Y124 samples.  The insets show the contact configuration for each sample.} \label{rawresfig}
\end{figure}

In contrast to the Y123 family, which has a single CuO chain with variable oxygen content, Y124 contains alternating
stacks of CuO$_2$ bilayers and {\it double} CuO chains that are stable and fully loaded. Single crystalline samples
(typical dimensions 400 x 80 x 30 $\mu$m$^3$) were flux grown in Y$_2$O$_3$ crucibles in a partial oxygen pressure of
400 bar \cite{AdachiNTNTHI98}. Magnetoresistance (MR) measurements were performed in the LNCMP pulsed field facility in
Toulouse. Although several samples were measured, clear SdH signals were seen in only two samples with the highest
$T_c$ values ($>$ 80K) and the largest MR response. Four-point resistance measurements were taken at a frequency of
40-80kHz with the current (10-20mA) applied {\bf I}$\|a$ (sample $\#$1, $T_c \sim$ 81K) and {\bf I}$\|b$ (sample $\#$2,
$T_c \sim$ 82K). The contact geometry for each crystal is shown in the insets of Fig.\ \ref{rawresfig}. The voltage
across the samples was recorded using an analogue to digital converter allowing the data to be post-processed. The
samples were cooled with either a dilution fridge ($\#$1) or a pumped $^4$He cryostat ($\#$2) housed in two different
magnet coils. In both cases, the samples were in direct contact with the cryogenic liquid at all temperatures. The
magnet coils had peak fields of 58T and 61T respectively and in the field range of interest here (45-61T), the maximum
field rise/fall rates were $\sim 2000$ T/s. Data for up and down sweeps showed negligible hysteresis.

The contact configuration of sample $\#$2 ({\bf I}$\|b$) allowed us to measure both the longitudinal ($\rho_{yy}$) and
Hall ($\rho^{\rm H}_{xy}$) resistivities by reversing the direction of the applied field, whereas for sample $\#1$
({\bf I}$\|a$),  $\rho_{xx}$ was measured directly. In order to improve the signal to noise ratio, data from several
(typically 2-3) pulses were averaged. Resistivity versus field data for both samples are shown in Fig.\
\ref{rawresfig}. The resistivities rise rapidly above the irreversibility field $H_{\rm irr}$ (which is slightly
different in the two samples) then increase more slowly. Above $H_{\rm irr}$, the resistivity is highly anisotropic
within the plane due to the presence of the highly conducting double-chain unit, as found in zero-field above $T_c$
\cite{Hussey97}. As in Y123-II \cite{Doiron-leyraudPLLBLBHT07}, $\rho^{\rm H}_{xy}$ is negative and varies linearly
with $H$, extrapolating to the origin. This linear field dependence suggests that the negative $\rho_{xy}$ is a normal
state effect, perhaps resulting from a Fermi surface reconstruction \cite{LeBoeuf} as discussed below, rather than a
mixed state effect. For both samples and in both the longitudinal {\it and} transverse components, small period (in
inverse field) SdH oscillations are visible which are a few percent of the total signal at maximum field.

\begin{figure}
\includegraphics*[width=0.8\linewidth,clip]{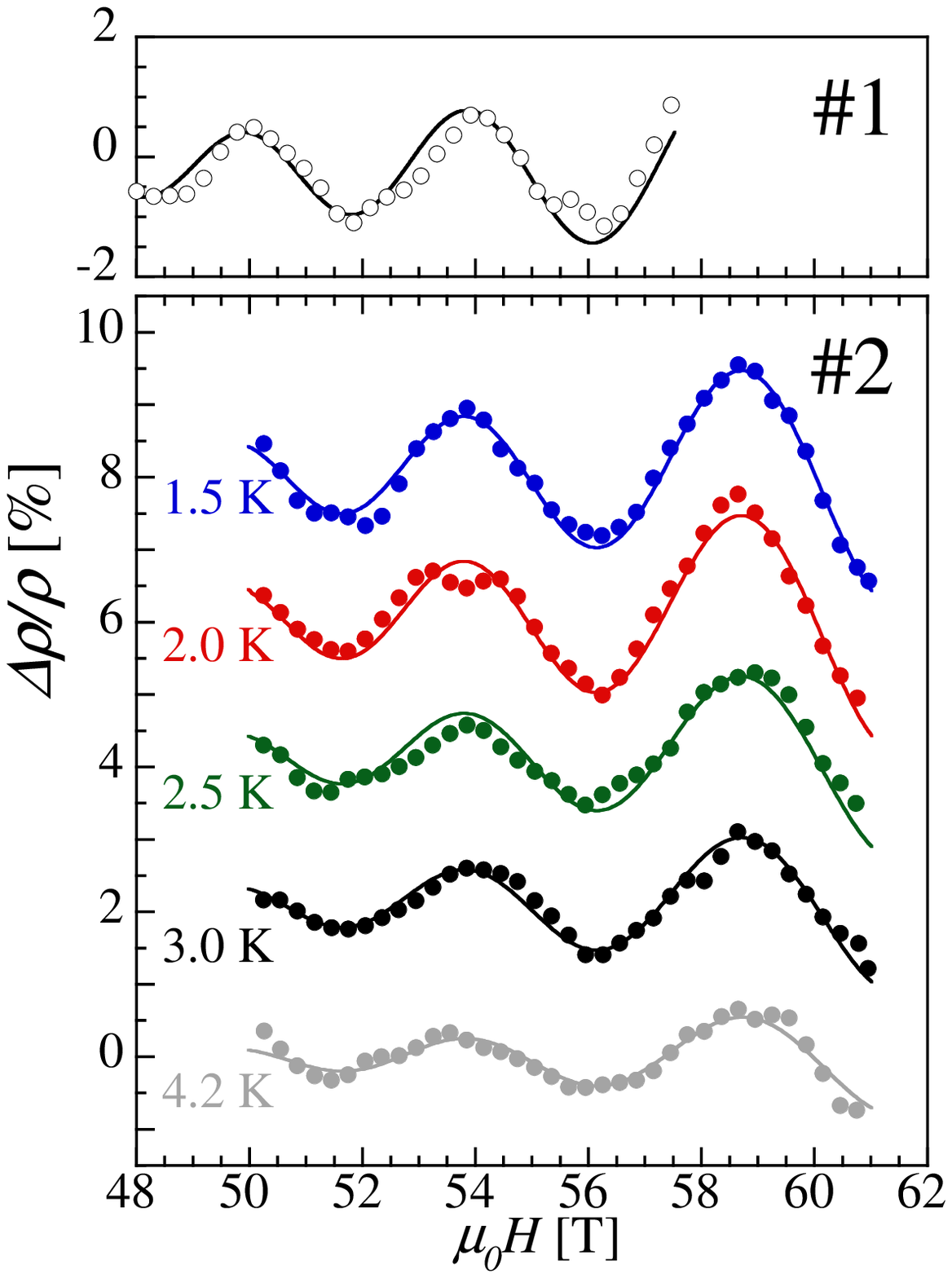}
\caption{(Color online). The oscillatory component of the magnetoresistance [$\Delta \rho/\rho(H)$] for both samples at
various temperatures. For \#1, $\rho$ = $\rho_{xx}(H)$ whilst for \#2, $\rho$ is a mixture of $\rho_{yy}(H)$ and
$\rho^{\rm H}_{xy}(H)$. In each case, a linear background has been subtracted and the data offset for clarity. The
solid lines are fits to Eq.\ (\ref{fiteq}) giving $m^* = 2.7\pm0.3m_e$.} \label{tdepsdhfig}
\end{figure}

In Fig.\ \ref{tdepsdhfig} we show the oscillatory component of the MR signal for both crystals at various temperatures.
Due to small differences in noise level for the two field directions, our study of the $T$-dependence of the
oscillations in sample \#2 was carried out with the field in one direction only. The resistivity $\rho$ shown in Fig.
\ref{tdepsdhfig} for \#2 is therefore a mixture of $\rho_{yy}$ and $\rho^{\rm H}_{xy}$.

We analyzed the data using the standard theory for SdH oscillations in a three-dimensional metal \cite{Richards73}, which is
essentially the same as the Lifshitz-Kosevich  model for the de Haas-van Alphen oscillations in the magnetisation
\cite{Shoenberg}
\begin{equation}
\frac{\Delta \rho}{\rho}=A H^\frac{1}{2} R_T R_D \sin\left(\frac{2\pi F}{B}+\phi\right)
\label{fiteq}
\end{equation}
where $A$ is a constant. The temperature damping factor $R_T = X/(\sinh X)$ with $X = (2\pi^2k_{_B}/\hbar e) (m^*T/B)$
and $m^*$ is the quasi-particle effective mass. The Dingle factor $R_D=\exp(-\pi/\omega_c\tau)=\exp[-(\pi \hbar
k_F)/(eB\ell)]$ describes the increase in damping as the orbitally averaged mean-free-path $\ell$ decreases or as the
average Fermi wavevector ($k_F$) of the orbit grows.

A fit of this equation to the data in Fig.\ \ref{tdepsdhfig} gives a single SdH frequency $F$ of $660\pm30$~T  for both
samples. Using the Onsager relation, $F=(\hbar/2\pi e) \mathcal{A}$, we find that $\mathcal{A}$, the cross-sectional
area of this orbit, corresponds to 2.4\% of the average cross-sectional area of the first Brillouin zone (27.9kT for
{\bf H}$\|c$). Assuming that the Fermi surface or pocket corresponding to this orbit is two-dimensional, and that there
are $n$ such pockets per CuO$_2$ sheet, Luttinger's theorem gives $p = nab\mathcal{A}/4\pi^2=0.024(2)n$ (where $a,b$
are the lattice constants). More importantly, $\mathcal{A}$ is found to be $1.25\pm 0.07$ larger than that observed in
Y123-II for which $p \sim 0.10$. Although it is difficult to pinpoint precisely the actual number of doped holes $p$
per CuO$_2$ plane in Y124, taking the average of $p$ from three independent estimates (by comparing $T_c$
\cite{LiangBH06} and the pseudogap temperature $T^*$ \cite{Hussey97, Takenaka94, Segawa01} with Y123 and the Seebeck
coefficient \cite{TallonBSHJ95, ZhouGD98}), one obtains $p \sim 0.14$. Thus $\mathcal{A}$ is found to scale
approximately linearly with the nominal hole densities of Y123-II and Y124.

\begin{figure}
\includegraphics*[width=0.7\linewidth,clip]{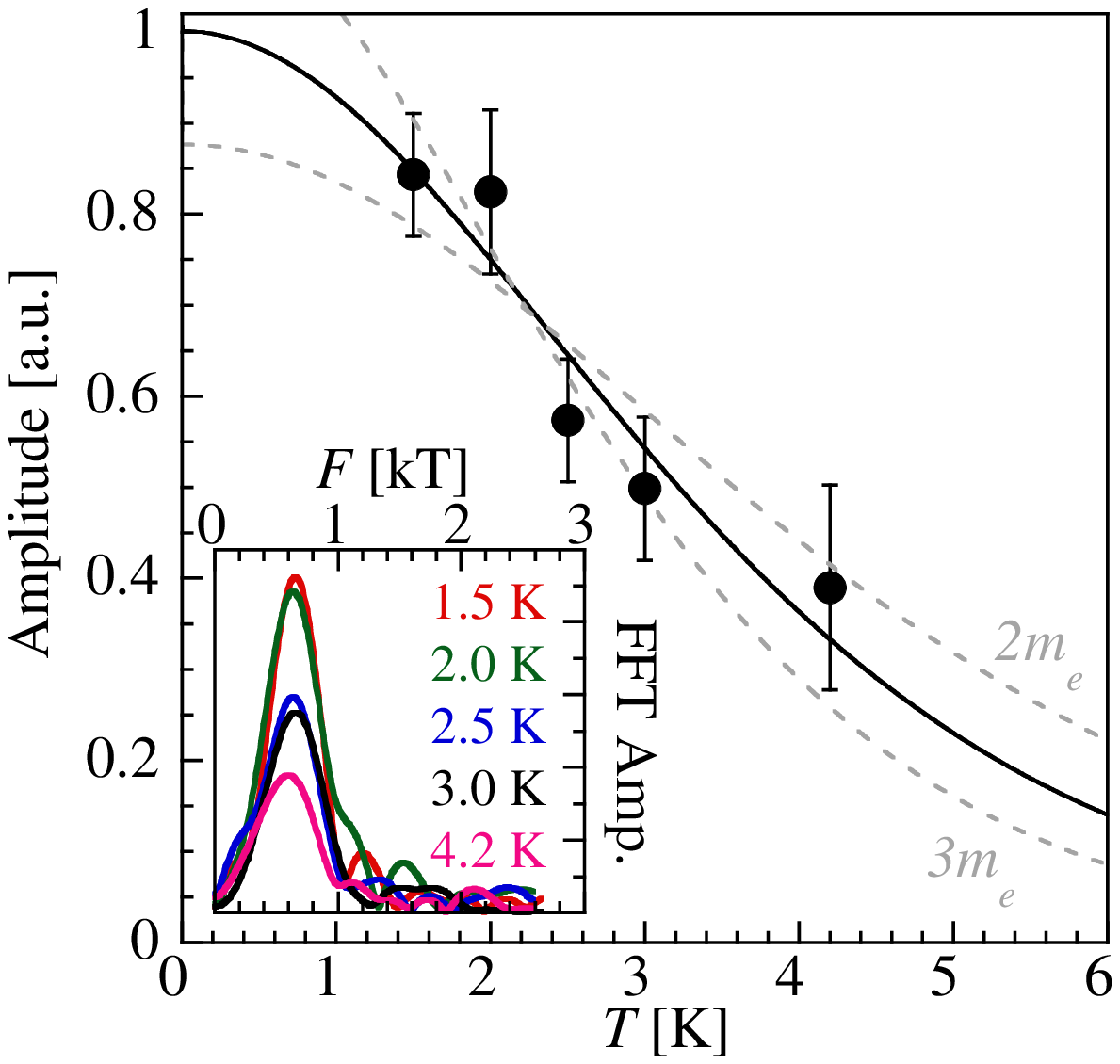}
\caption{(Color online). SdH signal amplitude versus temperature for sample \#2.  The solid line is a fit to the
standard expression for the temperature damping factor, $R_T$, yielding an effective mass of $2.5\pm0.3m_e$. The faint
dashed lines give the approximate range of masses consistent with the present data. Inset: Fast Fourier transform of
$\rho(1/B,T)$.} \label{massfig}
\end{figure}

As a first approach for determining $m^*$, we neglect the field dependence of  $R_T$ and fix $\omega_c\tau/B$, $F$ and
$\phi$ to the values found at the lowest temperature so that changes in the values of $A$ represent $R_T$ at an
effective field $B_{e}=55~T$. A fit of $R_T$ to $A(T)$ is shown in Fig.\ \ref{massfig} and gives $m^*=2.5\pm0.3m_e$,
where $m_e$ is the free electron mass. This approach slightly underestimates the mass, and a better approach is to fit
all the data at all temperatures to Eq.\ (\ref{fiteq}) simultaneously.   These fits, shown by solid lines in Fig.\
\ref{tdepsdhfig}, give $m^*=2.7\pm0.3m_e$. This is a heavier mass than that observed in Y123-II by a factor $1.4 \pm
0.2$.

Fast Fourier transforms of $\rho(1/B,T)$ are shown in the inset to Fig.\ \ref{massfig}. Both the frequency of the main
peak and the mass derived from the $T$-dependent amplitude are consistent with the direct fitting described above,
although the former technique is more accurate when so few oscillations are present.

From our fits in Fig.\ \ref{tdepsdhfig}, we obtain a Dingle factor ($e^{-\alpha/B}$, $\alpha=340 \pm 100~$T$^{-1}$) for
sample \#2 which for cylindrical pockets gives an average SdH mean-free-path $\ell_{\rm SdH}$$\simeq$ 90 $\pm$ 30\AA.
Unfortunately, we cannot obtain an independent estimate of $\ell_{\rm tr}$, the transport mean-free-path of the
in-plane carriers, as we do not yet know how many pockets are present. For sample \#1, the magnitude of the zero-field
residual in-plane resistivity $\rho_{a0}$ $\simeq$ 6 $\pm$ 2 $\mu\Omega$cm, estimated by fitting the high-field data to
the expression $\rho_a$($H$) = $\rho_{a0}$ + $AH^2$/(1+$BH^2$) that describes the form of the MR beyond the weak-field
limit \cite{Tyler98}. Assuming there are $n$ pockets per CuO$_2$ plane (and 2$n$ in total), this would give $\ell_{\rm
tr} = hc/2nk_Fe^2\rho_{a0}$ = 2000/$n$ \AA~($\pm$30$\%$). Here, $c$ = 13.6\AA~is the $c$-axis lattice spacing and $k_F$
= 1.4 nm$^{-1}$, as determined from the SdH frequency.

The above analysis implies that there is either a large number of Fermi pockets in Y124, or that the Dingle factor
grossly underestimates the transport lifetime \cite{footnote}. The latter could be attributed either to the stronger
effect of small-angle scattering on the dephasing of quantum oscillations, as seen in simple metals \cite{Richards73},
or to additional damping in the superconducting mixed state. Quantum oscillations have been observed in the
superconducting state of a number of different materials (see for example, Refs.\ \cite{JanssenHHMSW98,YasuiK02}).
Whilst the frequencies and effective masses are found to be the same as in the normal state, an additional damping is
observed which is related to the magnitude of the field dependent energy gap. We note however, that if the orbits
observed here originated from the nodal regions of the Fermi surface, as suggested by ARPES \cite{DamascelliHS03}, then
it may be expected that this additional damping is small because of the $d$-wave symmetry of the superconducting gap
function.

As mentioned above, conventional DFT band-structure calculations \cite{CarringtonY07,ElfimovSD07} for Y123-II show that
with small rigid shifts of bands ($\Delta E\sim\pm 20$ meV), a flat band arising from the CuO chain layer and apical
oxygen (BaO layer) can break through $E_F$ giving rise to small pockets, with frequencies rising up to $\sim$ 500~T and
bare masses $\sim 1-3$ $m_e$.  To investigate the possibility of similar pockets arising in Y124 we calculated the
band-structure of Y124 using the Wien2K package \cite{wien2k}, which is an implementation of a full-potential,
augmented plane wave plus local orbital scheme. The crystal structure from Ref.\ \cite{LightfootPJYMIK91}, and $10^4$
$k$-points in the full Brillouin zone were used for the consistency cycle.

\begin{figure}
\includegraphics*[width=0.85\linewidth,clip]{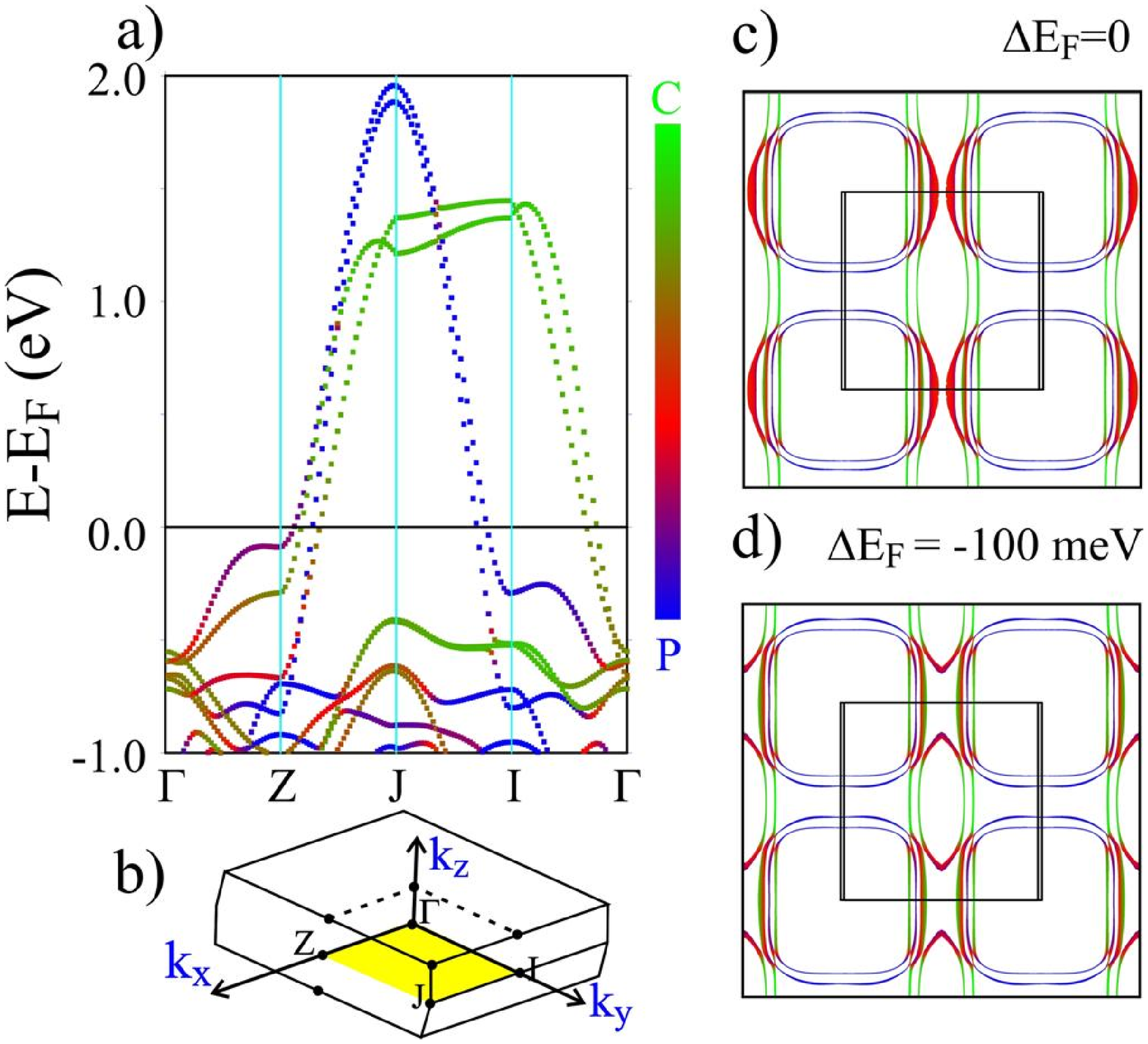}
\caption{(Color online). a) Band-structure of Y124 along symmetry directions in the basal plane. The color scale
reflects the band character: C=CuO chain character (including apical oxygen) P=CuO$_2$ plane character. b) Brillouin
zone of the Ammm space group of Y124, showing the symmetry points labelled in a). c) Fermi surface of Y124 viewed along
the $c$-axis. The shaded region is the first Brillouin zone. d) As for c) but with $E_F$ shifted down by 100~meV.}
\label{bsfig}
\end{figure}

The band-structure and Fermi surface are shown in Fig.\ \ref{bsfig} and are essentially the same as those reported
previously \cite{YuPF91,AmbroschdraxlBS91}. The Fermi surface consists of two large hole-like quasi-2D cylinders
centered on the zone corner (J) with mostly plane character and two warped quasi-1D sheets with mostly chain character.
Crucially, the position of the above mentioned CuO/BaO band is now $\sim$ 400 meV below $E_F$ (at the J point), and so
a very large shift of the energy of this band would be needed to produce Fermi surface pockets of similar origin to
those calculated for Y123-II.  In fact, $E_F$ needs to be shifted down by $\sim 80$meV before {\it any} new closed
orbits are formed for {\bf H}$\|c$. The orbit formed is centered around the zone center ($\Gamma$) (see Fig.\
\ref{bsfig}d) and has a high frequency ($\sim$3500~T for $\Delta E_F = -100$ meV) so cannot feasibly be responsible for
our observations. Given the similarity of our SdH results for Y124 and those for Y123-II, it therefore seems unlikely
that the oscillations in either material can be explained by conventional band-structure theory.

As proposed for Y123-II \cite{Doiron-leyraudPLLBLBHT07}, an alternative to the conventional band-structure approach is
to associate our SdH orbits with Fermi surface reconstruction, due, e.g. to spin-density-wave \cite{ChenYRZ07,
LinMillis05} or $d$-density-wave \cite{Chakravarty01} formation. Unfortunately, it is not possible at this stage to
have a complete picture of the (reconstructed) Fermi surface until we know precisely how many SdH orbits are present in
Y124. In this regard, measurements to higher fields should prove very informative.

In summary, we have observed Shubnikov-de Haas oscillations in the underdoped cuprate Y124. The frequency of the
oscillations is comparable with those observed in Y123-II, scaling approximately by the ratio of hole doping levels in
the two compounds. It is likely that these oscillations are a generic feature of underdoped cuprates perhaps resulting
from a reconstruction of the CuO$_2$ plane Fermi surface, although as both materials in which SdH oscillations have
been observed so far (Y123-II and Y124) contain quasi-1D CuO chain structures, an alternative origin cannot be
completely ruled out at this time.  Nevertheless our results provide precise information on how the pocket area and
effective masses vary as a function of doping, providing a strong constraint on theories which seek to explain their
electronic structure.

We thank E.~A.~Yelland and J.R. Cooper for showing us their data prior to publication \cite{Yelland07}, and
G.~G.~Lonzarich, J.W. Loram, and S.~M.~Hayden for useful discussions. We also thank EAY for help with the
band-structure calculations. This work was supported by the EPSRC (U.K.), NSERC (Canada), CIAR (Canada) and by
Euromagnet (EU contract 506239).

\end{document}